\documentclass[prb,aps,floats,showpacs,11pt,tightenlines,twocolumn,
	floatfix]{revtex4}

\usepackage{graphicx,color}

\newcommand{\be}{\begin{equation}}
\newcommand{\ee}{\end{equation}}
\newcommand{\bea}{\begin{eqnarray}}
\newcommand{\eea}{\end{eqnarray}}
\newcommand{\nn}{\nonumber}
\newcommand{\bm}[1]{\mbox{\boldmath${#1}$}}
\newcommand{\rr}{{\bf r}}

\newcommand{\vv}{{\bf v}}

\newcommand{\bwt}{\begin{widetext}}
\newcommand{\ewt}{\end{widetext}}

\begin{document}

\flushbottom


\title{Phase Transitions in Hexane Monolayers Physisorbed onto Graphite} 

\author{M.W. Roth}
\affiliation{\vspace{-.2cm} Department of Physics, University of
Northern Iowa, Cedar Falls, Iowa 50614}  

\author{C.L. Pint} 
\affiliation{\vspace{-.2cm} Department of Physics, University of
Northern Iowa, Cedar Falls, Iowa 50614}  

\author{Carlos Wexler}
\affiliation{\vspace{-.2cm} Department of Physics and Astronomy,
University of Missouri--Columbia, Columbia, Missouri 65211 \medskip}

\date{\today}

\begin{abstract}
We report the results of molecular dynamics (MD) simulations of a
complete monolayer of hexane physisorbed onto the basal plane of
graphite.  At low temperatures the system forms a herringbone solid. 
With increasing temperature, a solid to nematic liquid crystal
transition takes place at $T_1 = 138 \pm 2$K followed by another 
transition at $T_2 = 176 \pm 3$K into an isotropic fluid.  
We characterize the different phases by calculating various order
parameters, coordinate distributions, energetics, spreading pressure and
correlation  functions, most of which are in reasonable agreement with 
available experimental evidence.  In addition, we perform simulations 
where the Lennard-Jones interaction strength, corrugation potential strength
and dihedral rigidity are varied in order to better characterize the nature 
of the two transitions through.  We find that both phase transitions are
facilitated by a ``footprint reduction'' of the molecules via tilting, and to 
a lesser degree via creation of gauche defects in the molecules.
\end{abstract}

	
\maketitle

\section{Introduction}
\label{sec:introduction}
\vspace{-.2cm}

The physical adsorption (physisorption) of atoms and molecules onto a
substrate to form a quasi two-dimensional (2D) film has resulted in
the observation of a rich variety of behavior not realized in 
the system's corresponding three-dimensional (3D) or ``bulk''
state.  

Among the species used as adsorbates, alkanes are of considerable
interest because of their technological importance (e.g. as
lubricants) and because they are among the simplest families of
molecules of compounds whose members differ mainly in their length.
In fact, straight-chained n-alkanes represent a fine balance between
complexity (i.e. many internal degrees of freedom which also affect
the interaction {\em between} molecules) and the simplicity of their
structure (as compared to other organic molecules).  In fact, over the
past decades there has been a constant and renewed effort to better
understand the adsorption of alkaknes over a variety of substrates.   
Among the substrates used for physisorption studies, graphite, with
its excellent mechanical stability, availability and symmetry, has
proven to be one of the best choices.  A considerable amount of 
experimental, theoretical and computational work has, thus, been
devoted to these studies. 

Hexane on graphite was first studied by Krim {\em et al.} 
\cite{krim85} using LEED as well as neutron diffraction. For
submonolayer coverage a uniaxial incommensurate (UI) herringbone phase
is observed below ca. $T=151$ K, where a first-order melting
transition is found (but note that but a determination of the
molecular orientations was not possible). At low temperatures, as the
coverage is increased, the UI phase evolves continuously into a
$2\times 4 \sqrt{3}$ commensurate structure at completion.   
More recently, Taub and co-workers \cite{newtonthesis} completed extensive
neutron \cite{newtonthesis,taub88} and X-ray \cite{newtonthesis} diffraction
studies of hexane on graphite for submonolayer \cite{newtonthesis}, monolayer
\cite{newtonthesis,taub88,flemming92,flemming93,herwig97} and
multilater \cite{newtonthesis} coverages. Their findings indicate that
at low temperatures a complete monolayer forms a commensurate
herringbone structure which evolves with increasing temperature into a
rectangular centered solid/liquid coexistence region by ca.\ 150 K and
melts at around 175 K
\cite{newtonthesis,taub88,flemming92,flemming93,herwig97}.  
For submonolayer coverages, these authors proposed a structure
corresponding to a UI phase comprised of commensurate regions
separated by low-density fluid filled domain walls \cite{newtonthesis}.  

In the 1990's, Hansen {\em et al.} reported studies combining MD
computer simulations with neutron and x-ray diffraction studies
of butane and hexane on graphite at monolayer completion
\cite{flemming92,flemming93}.  Experimentally, butane melts abruptly
at around $T=116$ K directly from a solid commensurate
rectangular-centered herringbone (HB) phase into a
liquid. Orientational ordering about the surface normal is lost
through rotation about the center of mass, which is simultaneous with
melting. Hexane, on the other hand, undergoes a loss of translational
order at about 150 K into a phase with short-range order which is
thought to involve mobile rectangular-centered (RC) islands within a 
fluid. Then, at around $T=175$ K the system melts into a fluid. 

The computer simulations of Hansen {\em et al.}
\cite{flemming92,flemming93} showed that, for hexane, the internal
formation of {\em gauche} defects coupled with out-of-plane tilting is
concurrent with melting. The necessary creation of in-plane room by
either gauche defects or tilting at the onset of melting is referred
to as a ``footprint reduction'' mechanism.  These simulations showed,
furthermore, that the nature and temperature of the transition is very
sensitive to the presence of gauche defects, as the melting
temperature raises drastically when gauche defect formation is
suppressed.  However, the resulting melting temperature of a hexane
monolayer was found to be  222 K, significantly different from the
experimental value of 175 K. This discrepancy was addressed in a
series of MD simulations of hexane on graphite by Velasco and Peters
\cite{velasco95} which found a  better agreement with experimental
values of the melting temperature when a considerably lower
adsorbate-interaction strength was used, and under those conditions
molecular gauche defect formation becomes almost irrelevant to melting.
(In this paper we find the presence of an intermediate nematic
phase.  In the herringbone solid to nematic phase transition we find
gauche defects to be irrelevant.  For the nematic to isotropic liquid
phase transition, we find that both mechanisms contribute to the
``footprint reduction'', although it is apparent that tilting is 
the dominant effect.)

More recently, Peters and coworkers have also examined the behavior of
monolayer \cite{peters96a} and bilayer \cite{peters96b} hexane on
graphite using MD simulation techniques.  The monolayer study compared
MD results based on the standard (isotropic) unified atom (UA)
approximation to simulations where an {\em anisotropic} force field
was employed.  Using a computational cell of 112 hexane molecules, and
performing MD simulations corresponding to 200,000 steps of 1 fs both
models yielded the same basic physics, especially in regards to the
characterization of the phases observed (the transition temperature
from commensurate herringbone to the rectangular centered
orientationally ordered phase is found at around 150 K and the melting
temperature at 175 K in both cases, the main difference between the
two models is that the anisotropic model promotes more in-plane
mobility from vacancy creation due to increased molecule tilting at
all temperatures).

Even though hexane on graphite has been studied quite well
experimentally, it has not been as thoroughly investigated
computationally, and there is still some doubt about the roles played
in the phase transitions by the internal degrees of freedom of the
molecules (mainly gauche defects) and their relative importance as the
adatom-substrate interaction is varied.  Furthermore, it is not
self-evident that the simulations have achieved ergodicity to a
satisfactory degree, as only a relatively limited simulation time has
been explored so far (maximum of 2 ns in reference
\onlinecite{peters96a}).  The general purpose of the work reported
here is to enrich our understanding of phase transitions in
physisorbed molecular systems, in particular of hexane in
graphite. Specifically, we wish to:    
{\em (i)} more sharply delineate the phase transitions of
	hexane on graphite using longer simulations (up to 5 times
	longer than previously reported);   
{\em (ii)} to understand the effects of varying the graphite
	corrugation strength and the adatom-adatom interaction
	strength on the observed phase transitions;   
{\em (iii)} to further delineate the role of the internal degrees of
	freedom of the adsorbed molecules, i.e. the bond bending and
	dihedral torsion in the observed phase transitions, and   
{\em(iv)} to gain insight in the energetics and possible mechanisms
	for the observed phase transitions.  

Since the system under consideration is, to some degree, (quasi-)
two-dimensional, one may have expected that the observed phase
transitions would show some of the universal features expected from
the beautiful theory of 2D melting developed in the 1970's by
Kosterlitz, Thouless, Halperin, Nelson and Young
\cite{kosterlitz73,kosterlitz74,halperin78,nelson79,strandburg88}. 
However, the importance of the out-of-plane behavior of the 
hexane monolayer, and the ``polarizing'' effect of the substrate
corrugation make it difficult to reliably place these transitions in
a general theoretical context.  Furthermore, tt is evident that the
the relatively small system sizes achievable in MD simulations does
not permit a complete characterization of the phase transitions
observed (e.g. their order), but sometimes the modality of energy
distributions near the critical regions may give {\em indications} on
what behavior to expect.   
        
The paper is organized as follows. In Sec.\ \ref{sec:comp_details} we
discuss the computer simulation procedures.  The results of the simulations
are presented in Sec.\ \ref{sec:results}, and are discussed in detail
in Sec.\ \ref{sec:discussion}.

\vspace{-.1cm}
\section{Computational Details}
\label{sec:comp_details}
\vspace{-.2cm}

\begin{figure*}
\begin{center}
\leavevmode
\includegraphics[width=6in]{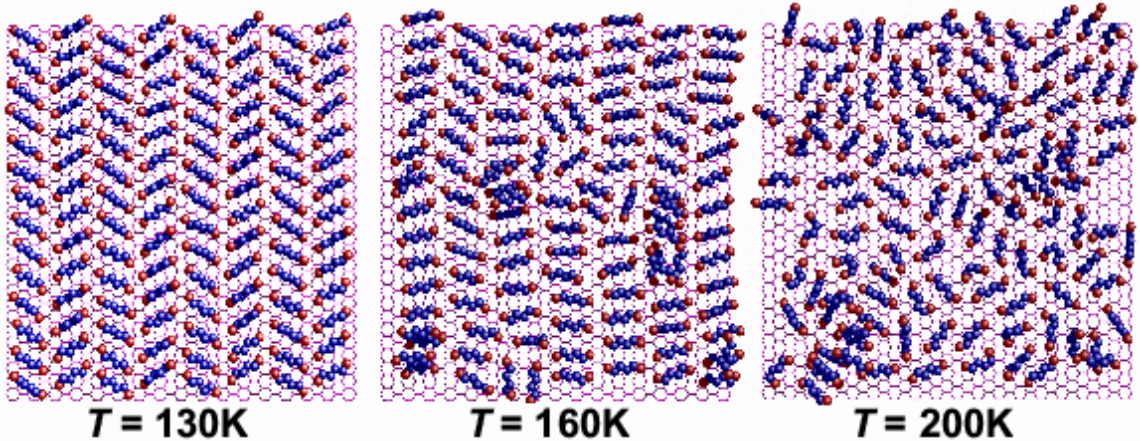}
\vspace{-0.5cm}
\end{center}
\caption{\label{fig:topview} 
Simulation cell showing snapshots of typical (equilibrium)
configurations for the hexane/graphite system in the commensurate
herringbone solid ($T=130$ K), the orientationally ordered nematic
($T=160$ K) and isotropic liquid ($T=200$ K) phases. CH$_3$
end-of-chain pseudoatoms are red, CH$_2$ are yellow and the C-C bonds
in the graphite substrate are purple (the graphite lattice spacing is
$a = 2.46$ {\AA}). The computational cell has dimensions
$m\times4\sqrt{3}a\simeq68.1735$ {\AA} and $n\times2a=68.88$ {\AA} in
the $x$ and $y$ directions respectively, where $m=4$ and $n=14$, and
periodic boundary conditions are used in $x$ and $y$ directions.
There are $2\times m \times n = 112$ hexane molecules. } 
\end{figure*}

The hexane molecules in our study are modeled with two methyl (CH$_3$)
pseudo atoms on its ends and four methylene (CH$_2$) pseudo atoms in 
between. The lowest energy  conformation of the molecule consists of
all pseudo atoms being co-planar, forming a ziz-zag pattern with
114$^\circ$ three-body bond angles (dihedral angles) and bond lengths
(nearest neighbor separations) of 1.54 {\AA}.  In the simulations that
follow we keep the the bond lengths fixed, but allow the other
internal degrees of freedom to vary (bond angle bending and dihedral angle 
torsion).  The graphite substrate modeled as being static, containing
an infinite number of stacked graphene sheets (see
Fig. \ref{fig:topview}).  The  structure contains point carbon atoms
at the vertices of all bonds and the bond length is taken to be 2.46
{\AA}. 

\vspace{-.1cm}
\subsection{Method}
\vspace{-.2cm}

A constant particle number, planar density and temperature
($N=672,\ \rho=1,\ T$) molecular dynamics (MD) method is employed to
conduct atomistic simulations a system of $N_m = 2\times m \times n =
112$ hexane molecules at monolayer completion ($m=4$ and $n=14$). The
rectangular computational cell (see Figure \ref{fig:topview}) has its
corners coincident with graphite hexagon centers and has dimensions
$m\times4\sqrt{3}a\simeq68.1735$ {\AA} and $n\times2a=68.88$ {\AA} in
the $x$ and $y$ directions respectively.  Such a cell gives a planar
density of  $\rho=0.02385$ molecules/{\AA}$^2$ and an area per
molecule of 42 {\AA}$^2$. Each hexane molecule consists of six united
atom (UA) pseudoatoms--two methyl (CH$_3$) and four methylene
(CH$_2$). Periodic boundary conditions are utilized for each
pseudoatom in the $x,y$ directions and free boundary conditions are
present in the vertical $z$ direction. To perform the simulations at
constant temperature, we periodically rescale the velocities so that
equipartition is satisfied for the center-of-mass, rotational and
internal temperatures: 
\bea
\label{eq:temperatures}
	T_{CM} &=& \frac{1}{3 N_m k_B} 
        \sum_{i=1}^{N_m} M_i v_{i,CM}^2 \,, \nn \\
	T_{ROT} &=& \frac{1}{3 N_m k_B} 
\sum_{i=1}^{N_m}{\bm{\omega}}_i^T  
        {\cal I}_i {\bm{\omega}}_i \,, \\
	T_{INT} &=& \frac{1}{(2n_C-5)N_m k_B} \sum_{i=1}^{N_m}
\sum_{j=1}^{n_C} m_{ij} \nn \\ 
&& \hspace{-.3cm} \times [(\vv_{ij} - \vv_{i,CM}) 
 - {\bm{\omega}}_i
 	\times (\rr_{ij}-\rr_{i,CM})]^2 , \nn
\eea
respectively (no significant differences were found when
``thermalizing'' a subset of the temperatures above).  Here $N_m$ is
the number of molecules, $n_C$ the number of pseudoatoms per molecule,
the $i$ index runs over the molecules ($1\le i\le N_m$) and the $j$
index runs over the pseudoatoms within a given molecule ($1\le j\le
n_C$).  Furthermore, for each molecule, $M_i$ is the mass,
${\bf{v}}_{i,CM}$ is the center-of-mass velocity, ${\bm{\omega}}_i$ is
the angular velocity and ${\cal I}_i$ is the moment of inertia tensor
(relative to the COM).  Last, ${\bf{v}}_{ij}$ is the velocity of the
$j$-th pseudoatom in the $i$-th molecule, and $k_B$ is Boltzmann's
constant. 

The time step for all simulations is chosen to be 1 fs and integration
of the equations of motions is achieved  with a velocity Verlet RATTLE
\cite{allen88} algorithm which keeps the pseudoatom bond lengths
constant.  Runs are typically started from a herringbone
configuration, as suggested by experimental evidence (e.g. the neutron
diffraction data of Ref.\ \onlinecite{krim85}), and allowed to
equilibrate for roughly 3--5$\times 10^5$ steps, averages are taken
for 0.5--1.0$\times 10^6$ steps after equilibration.  Some runs are
started from the final configuration of some lower temperature run.
A good measure of convergence is that incorporating the results from
both types of initial conditions does not create noticeable scatter in
the results.  The only hysteresis observed is that, as is usual in
these type of simulations, ``freezing'' by starting from a high
temperature phase and lowering the temperature is not readily
achieved. 

\vspace{-.0cm}
\subsection{Interaction Potentials}
\vspace{-.0cm}

The model adopted for our study of adsorbed hexane molecules on
graphite corresponds to having both non-bonded and bonded
interactions.  The first of two non-bonded interactions is the
adsorbate-adsorbate interaction between pseudoatom atom $i$  and $j$
separated by a distance $r_{ij}$, which is modeled by a Lennard-Jones
pair potential  
\be
\label{eq:lj}
        u_{LJ}(r_{ij}) = 4 \epsilon_{ij} \left[ 
        \left( \frac{\sigma_{ij}}{ r_{ij}} \right)^{12} 
        - \left( \frac{\sigma_{ij}}{ r_{ij}} \right)^{6} \right] \,.
\ee
Lorentz-Bertholot combining rules
\be
\label{eq:lb}
	\sigma_{ij} = \frac{\sigma_{i} + \sigma_{j}}{2} \,, 
        \hspace{1cm}
        \epsilon_{ij} = \sqrt{\epsilon_i \epsilon_j} \,,
\ee
are applied in order to describe mixed interactions when particles $i$
and $j$ are of different types; lattice sums are taken out to 10 {\AA}
and values for the potential parameters used in our work
\cite{martin98} are given in Table \ref{tabl:non-bond}.

The second non-bonded interaction is the adatom-graphite
interaction. The graphite is modeled as being of infinite extent in
the $xy$ plane and semi-infinite in the vertical $-z$ direction. 
The adatom-graphite interaction potential, however, is not taken as a
discrete sum over the carbon atoms. Steele's expansion \cite{steele73}
is utilized because it exploits the fact that the symmetry of the
substrate breaks the adatom-graphite interaction into a vertically
(laterally averaged) portion and a much weaker portion related to the
undulation in the interaction due to the graphite hexagonal
symmetry. Such a representation saves a considerable amount of
computing time.  In other words, the interaction of the adsorbate
atoms with the substrate is the Fourier expansion
\be
\label{eq:steele1} 
  u_{gr}(\rr_i) = E_{0i}(z_i) + \sum_{n=1}^\infty 
	E_{ni}(z_i) f_{n}(x_i.y_i) \,,
\ee
with
\bwt
\bea
\label{eq:steele2} 
	E_{0i}(z_i) &=& \frac{2 \pi q \epsilon_{gr} \sigma_{gr}^6}{a_s} 
	\left[ \frac{2 \sigma_{gr}^6}{45 d (z_i + 0.72 d)^9} +
		\frac{2 \sigma_{gr}^6}{5 z_i^{10}} - \frac{1}{z_i^4} -
		\frac{2 z_i^2 + 7 z_i d + 7 d^2}{6 d (z_i+d)^5} \right] \,, 
\nn \\
	E_{ni}(z_i) &=& \frac{2 \pi \epsilon_{gr} \sigma_{gr}^6}{a_s}
	\left[ \left(\frac{\sigma_{gr}^6}{30}\right) 
		\left(\frac{g_n}{2 z_i}\right)^5 K_5(g_n z_i) -
	        2 \left(\frac{g_n}{2 z_i}\right)^2 K_2(g_n z_i) \right] \,, 
\\
	f_{1} (x_i,y_i) &=& 
2 \cos \Bigl[\frac{2 \pi}{a}(x + \frac{y}{\sqrt{3}})\Bigr]
+2 \cos\Bigl[\frac{2 \pi}{a}(x - \frac{y}{\sqrt{3}})\Bigr]
+2 \cos\Bigl[\frac{4 \pi}{a}(\frac{y}{\sqrt{3}})\Bigr] \,, \nn
\eea
\ewt
Here $g_n$ is the modulus of the $n$-th graphite reciprocal lattice
vector and the $K_n(x)$ are modified Bessel functions of the second
kind. The interaction is obtained by summing over an infinite number
of graphene sheets. Only $f_{1}(x_i,y_i)$ is defined above because the
sum in $x,y$-dependent part of equation (\ref{eq:steele1}) converges
extremely rapidly and only the $n=1$ term is needed. All potential
parameters for the adsorbate-substrate interaction are identified in
Table \ref{tabl:non-bond}.

\begin{table}[t]
\caption[]{Non-bonded potential parameters used in the simulations}
\begin{center}
\begin{tabular}{|c|c|}		
\toprule
{\bf Parameter} & {\bf Value} 
	\\ \hline
$\epsilon_{\rm CH_2}$, $\epsilon_{\rm CH_3}$ 	&	72 K 
\\ \hline
$\sigma_{\rm CH_2}$, $\sigma_{\rm CH_3}$  & 3.92 {\AA} \\ \hline
$q$	& 	2  					\\ \hline 
$a$	&	2.46 {\AA} 					\\ \hline
$a_s$	& 	5.24 {\AA}$^2$ 			\\ \hline
$d$	&	3.357 {\AA}					\\ \hline
$\epsilon_{gr}$		&	44.89 K		\\ \hline
$\sigma_{gr}$		&	3.66 {\AA}		\\ 
\botrule
\end{tabular}
\end{center}
\label{tabl:non-bond}
\end{table}

There are two types of bonded interactions considered in this work:
bond  angle bending and dihedral angle bending (we assume that the
bond lengths are fixed at 1.54 {\AA}, the RATTLE algorithm allows for
constrained solution of the equation of motion \cite{allen88}).  The
bond angles are assumed to be harmonic; their bending potential energy
comes from Martin and Siepmann \cite{martin98} and is given by   
\be
\label{eq:ubend}
u_{bend}(\theta_b) = \frac{k_\theta}{2} (\theta_b - \theta_0)^2 \,,
\ee
where $\theta_b$ is the bond angle, $\theta_b$ is the equilibrium bond
angle and $k_{\theta}$ is the angular stiffness. In addition, the
expression for the energy of dihedral bending (torsion)
\cite{peters96a,padilla91} is of the form 
\be
\label{eq:utors}
	u_{tors} (\phi_d) = \sum_{i=0}^5 c_i (\cos \phi_d)^i \,,
\ee
where $\phi_d$ is the dihedral angle and the $c_i$ are constants.  All
values for the bonded potential parameters are shown in Table
\ref{tabl:bond}.   This potential has one global minimum for the {\em
trans} configuration $\phi_d = 0$ and two local minima with $\sim$234
K higher in energy for the two {\em gauche} configurations at $\phi_d
= \pm 2 \pi/3$.  

The full potential energy can thus be written:
\bea
&& U = U_{\rm nb} + U_{\rm b} \,, \\
&& U_{\rm nb} = 
	\frac{1}{2}\!{\sum_{i,i'=1}^{N_m}}\!\!' \sum_{j,j'=1}^{n_C}\!\!
		u_{LJ}\!(|\rr_{ij}\!-\!\rr_{i'j'}|)
	+ \sum_{k=1}^{N} \! u_{gr}\!(\rr_k) , \nn \\
&& U_{\rm b} = \sum_{i=1}^{N_m} \Bigl[
	\sum_{j=1}^{n_C-2}\! u_{bend} 
	+ \sum_{j=1}^{n_C-3}\! u_{tors}
	+ \sum_{j=1}^{n_C-4}\! u_{LJ}  
	\Bigr] . \nn
\eea

\begin{table}[hb]
\caption[]{Bonded potential parameters used in the simulations}
\begin{center}
\begin{tabular}{|c|c|} 		
\toprule
{\bf Parameter} & {\bf Value} 	
	\\ \hline
$k_\theta$ 	&	62,793.6 K/rad$^2$ 		\\ \hline
$\theta_0$	&	114$^\circ$			\\ \hline
$c_0$		&	1,037.76 K			\\ \hline
$c_1$		&	2,426.07 K			\\ \hline
$c_2$		&	81.64 K			\\ \hline
$c_3$		&	-3,129.46 K			\\ \hline
$c_4$		&	-163.28 K			\\ \hline
$c_5$		&	 -252.73 K			\\	
\botrule
\end{tabular}
\end{center}
\label{tabl:bond}
\end{table}

\section{Results}
\label{sec:results}
\vspace{-.4cm}

We performed MD simulations as described above for many temperatures
in the 100--200 K range.  Figure \ref{fig:topview} shows snapshot of
typical configurations of the simulated hexane/graphite system for
three characteristic temperatures where the herringbone (HB)
commensurate solid, the ``nematic,'' and the fluid phases are
observed.  It should be mentioned that promotion of molecules to
higher layers was not very significant at any of the temperatures
studied. From this figure it is evident that in the low-temperature
phase molecules are arranged in a HB structure where the molecules'
azimuthal angles $\phi_i \in [0,\pi]$ \cite{directors} take
predominantly values near ${\{30^\circ, 150^\circ\!\}\,}$  
($\phi_i$ is defined as the angle that the axis of the smallest moment
of inertia for molecule $i$ makes with the $x$-axis).  Moreover, the
molecules are in registry with the substrate.  Clearly this HB phase
is solid. In the high-temperature regime, it is also quite clear that
the molecules are randomly positioned and oriented (and in particular
not in registry with the substrate), which is strongly suggestive of a
liquid phase.  The situation of the mesophase, however, is not as
evident.  It is clear that it has a definite orientational order,
quite different than the HB (most azimuthal angles centered around
0$^\circ$, the molecules appear not to be commensurate with the
substrate, and it is not clear whether the system is in a liquid
crystalline state (i.e. a {\em nematic}), or in a solid phase (i.e. a
rectangular centered solid).  The exact nature of the different phases
(and in particular of the mesophase) is addressed more quantitatively
below, by means of examination of various order parameters,
energy analysis, spreading pressure, distributions and correlation functions. 

\vspace{-0.6cm}
\subsection{Order parameters}
\label{sec:OPs}
\vspace{-.4cm}

\begin{figure}
\begin{center}
\leavevmode
\includegraphics[width=3.in]{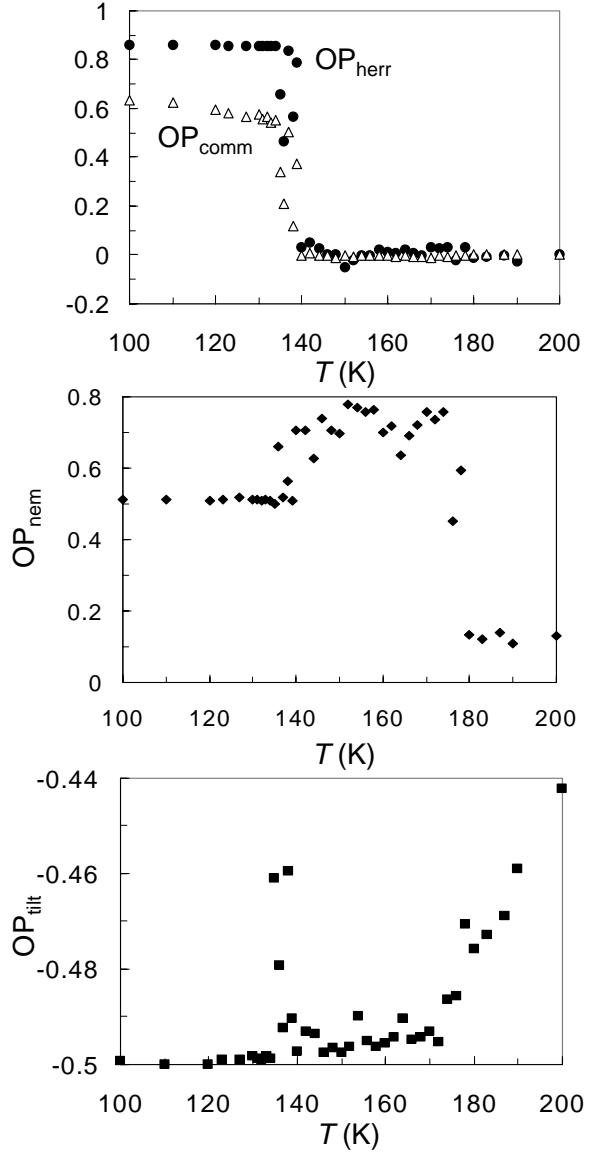}
\vspace{-.5cm}
\end{center}
\caption{\label{fig:OPs} 
	Structural order information. The following order parameters (OP) are
	presented as functions of temperature:  
	Herringbone $OP_{herr}$ (solid circles), center of mass
	$OP_{comm}$ (open triangles), nematic $OP_{nem}$ (solid diamonds) 
	and tilt $OP_{tilt}$ (solid squares) .}
\end{figure}

Figure \ref{fig:OPs} shows the temperature dependence of several
molecular order parameters which are useful in characterizing the
system's behavior through and across phases.  The herringbone order
parameter is used to quantify the in-plane orientational ordering of
the molecular axes and is defined as   
\be
\label{eq:OP_herr}
	{OP}_{herr} \equiv \frac{1}{N_m} \left<
		\sum_{i=1}^{N_m} (-1)^j  \, \sin (2 \phi_i) \right> \,,
\ee
where the sum is peformed over all molecules, the integer $j$ is
assigned so as to take the two sublattices into account and the
averages correspond to sampling a large number of uncorrelated
configurations in each MD simulation. For a ``perfect herringbone''
with $\phi_i \in \{45^\circ,135^\circ\}$, $OP_{herr} = 1$, obviously
$OP_{herr}$ vanishes if the two sublattices cannot be distinguished
(e.g. if all angles are clustered around {\em one} angle or if all
angles are sampled equally) either statically (over the lattice) or
dynamically (over time).  For the HB phase, where the $\phi_i$'s are
clustered around 30$^\circ$ and 150$^\circ$ (for further evidence see
the 130 K curve in Fig.\ \ref{fig:P_phi}), we find $OP_{herr} \simeq
\sqrt{3}/2 \simeq 0.866$, as observed for temperatures smaller than
$T_1 = 138 \pm 2$ K.  Not surprisingly $OP_{herr} \simeq 0$ for the
high temperature phase as essentially all azimuthal angles are equally
sampled (see the 200 K curve in Fig.\ \ref{fig:P_phi}).  We also find
$OP_{herr} \simeq 0$ for the mesophase, which is also evident from the
$\phi_i$ distributions which show a single significant orientation and
thus the lack of two angular distinct sublattices (see the 160 K curve
in Fig.\ \ref{fig:P_phi}). 

To investigate whether the adsorbed molecules are in registry with the
graphite substrate we also measure the commensurate order parameter
$OP_{comm}$, defined as   
\be
\label{eq:OP_comm}
	{OP}_{comm} \equiv \frac{1}{6 N_m} \left<
  \sum_{i=1}^{N_m}\sum_{s=1}^{6} e^{-i {\bf g}_s \cdot \rr_i} \right> \,, 
\ee
where the outer sum runs over all $N_m$ molecules at center-of-mass
position $\rr_i$, and the inner sum runs over all six graphite
reciprocal lattice vectors ${\bf g}_s \in \{ (\pm\pi/3a,0),$
$(\mp\pi/6a,\pm\pi/\sqrt{3}a),$ $(\pm\pi/6a,\pm\pi/\sqrt{3}a)\}$.   
$OP_{comm}$ takes on a value of unity were
all molecular centers to remain statically graphite hexagon centers
and vanishes in the limiting case of the molecular centers uniformly
sampling positions in the $xy$ plane.  As with $OP_{herr}$, Such a
sampling could occur either in a static fashion (an infinite
incommensurate solid) or in a dynamic fashion such as in a
non-registered film.  It is evident from Fig.\ \ref{fig:OPs}, that as
the HB phase disappears at $T_1 = 138 \pm 2$ K, the film also looses
registration with the substrate.  This is also evident in the decrease
of the corrugation energy (see Sec. \ref{sec:energy} and bottom panel
of Fig.\ \ref{fig:ulj_u1}). 

Both $OP_{herr}$ and $OP_{comm}$ are useful in providing information
on the solid to nematic transition at $T_1 = 138 \pm 2$ K but are
unable to provide any characterization of the nematic to isotropic
liquid transition.  The nematic order parameter $OP_{nem}$ provides
precisely the needed information and is defined by  
\be
\label{eq:OP_nem}
	{OP}_{nem} \equiv \frac{1}{N_m} \left<
		\sum_{i=1}^{N_m}  \cos 2 (\phi_i - \phi_{dir}) \right> \,,
\ee
where $\phi_{dir}$ is an ``average'' azimuthal direction defined by
maximizing the nematic order parameter: $\partial{OP}_{nem}/\partial
\phi_{dir} \equiv 0$: 
\be
\label{eq:phidir}
	\phi_{dir} = \frac{1}{2} \tan^{-1} \left[
	{
		\sum_{i=1}^{N_m} \sin 2 \phi_i} /
	{
	\sum_{i=1}^{N_m} \cos 2 \phi_i}
	\right]\,.
\ee
Note that due to the 1/2 pre-factor, it is important to use the
4-quadrant version of the $\tan^{-1}$ function to uniquely define
$\phi_{dir}$ (the 2-quadrant version leads to the possibility of
$\pi/2$ rotations which are relevant for the final value of $OP_{nem}$
in some cases).   

The center panel of Fig.\ \ref{fig:OPs} shows the temperature
dependence of the nematic order parameter.  This order parameter
clearly indicates that, in addition to the HB to nematic phase
transition at $T_1 = 138 \pm 2$ K, a second {\em orientational} phase
transition occurs at $T_2 = 176 \pm 3$ K.  As shown earlier for the
$OP_{herr}$, $OP_{nem}$ also is clearly correlated to the azimuthal
angle distributions (Fig.\ \ref{fig:P_phi}).  For a ``perfect
herringbone'' solid ($OP_{herr}=1$), $\phi_{dir}$ is completely {\em
undefined} and $OP_{nem}$ vanishes, but for the solid commensurate
herringbone phase of hexane on graphite ($OP_{herr}=\sqrt{3}/2$),
$\phi_{dir}=0$ and the limiting value of $OP_{nem}$ is 0.5.  In a
perfect nematic phase, all molecules share the same orientation and
$OP_{nem}=1$.  Our results of $\max[OP_{nem}] \simeq 0.8$ show the
presence of significant fluctuations in the molecular orientations and
of the presence of differently oriented domains, both features are
also clearly qualitatively visible in the 160 K data presented in the
configuration snapshot (Fig.\ \ref{fig:topview}) and azimuthal
distribution (Fig.\ \ref{fig:P_phi}).  In the liquid phase we would
expect $OP_{nem}$ to vanish, the small finite value in the high
temperature phase is an indication of a small residual order likely
due to the small system size. Note that contrary to the previous order
parameters, $OP_{nem}$ does not vanish due to dynamic fluctuations
leading to an overall rotation of the average director over time.  We
would like to remark that our results for the nematic order parameter
(center panel of Fig.\ \ref{fig:OPs}) differ from those found by
Peters and Tildesley (PT) \cite{peters96a}:  \emph{(a)} they find
$OP_{nem} \simeq - 0.5$ for the HB phase; \emph{(b)} they find a
relatively small $\max[OP_{nem}] \simeq$ 0.0--0.4; \emph{(c)} they
find the HB to nematic transition at a higher temperature (ca.\ 150
K); and \emph{(d)} they find a relatively undefined nematic to liquid
phase transition at ca.\ 175 K.  The difference in \emph{(a)} (and to
some extent \emph{(b)}) is due to a factor of 1/2 difference in the
definition of $\phi_{dir}$ in equation (\ref{eq:phidir}).  The higher
``importance'' of the nematic mesophase in our work [points
\emph{(b--d)}] we belive is due to the considerably better statistics
in our runs (whereas PT used $5\times 10^4$ thermalization and
$2\times10^5$ averaging steps, we typically ran our simulations for
3--5$\times 10^5$ thermalization and 0.5--1.0$\times 10^6$ averaging
steps).   

The tilt order parameter is designed to provide a measure of the
out-of-plane orientational order of the system-–-in particular, the
amount of molecular tilting---and is defined as
\bea
\label{eq:optilt}
	OP_{tilt} &\equiv&  
	\frac{1}{N_m} \left<\! \sum_{i=1}^{N_m}
	\langle P_2(\cos \theta_i) \right>  \\
	&= &\frac{1}{2N_m} \left<\! \sum_{i=1}^{N_m}
	(3 \cos^2 \!\theta_i -1) \!\!\right> \!, \nn
\eea
were $P_2(x)$ is the Legendre polynomial and $\theta_i$ is the angle
that the smallest moment of inertia axis of molecule $i$ makes with
the normal to the substrate.   Clearly $OP_{tilt}$ is closely
associated with the tilt angle distribution $P(\theta)$ (see Fig.\
\ref{fig:P_theta}). When all molecules are parallel to the
surface $OP_{tilt} = -0.5$, and $OP_{tilt}=1$ when all molecules are
perpendicular to the surface. The bottom panel of Fig.\ \ref{fig:OPs}
shows the temperature dependence of $OP_{tilt}$.  At low temperatures
the molecules clearly are (as expected) roughly parallel to the
graphite substrate to maximize their adsorption energy.  At the HB to
nematic transition, $OP_{tilt}$ has a significant peak, indicating a
tendency of the molecules to librate our of the surface plane, but
this tendency is reduced again as the nematic phase is better
``established.''  For increasing temperatures $OP_{tilt}$ barely
incrases until the nematic to liquid phase transition, where it starts
a significantly faster growth.  
Tilting of the molecules can obvioulsy be a significant factor in
the phase transitions, as the ``footprint'' of the molecules is
reduced. To put things in perspective, however, $OP_{tilt}(200 \; {\rm
K}) \simeq -0.44$ roughly corresponds to an average libration of only
11$^\circ$ out of the plane. Nevertheless, for a full monolayer
coverage the resulting ``footprint reduction'' can be
significant. (For more comments on the stiffling of tilting by the
substrate potential see Sec.\ \ref{sec:variations}.)

\vspace{-0.2cm}
\subsection{Energy and Spreading Pressure}
\label{sec:energy}
\vspace{-.3cm}

\begin{figure}
\begin{center}
\leavevmode
\includegraphics[width=3.in]{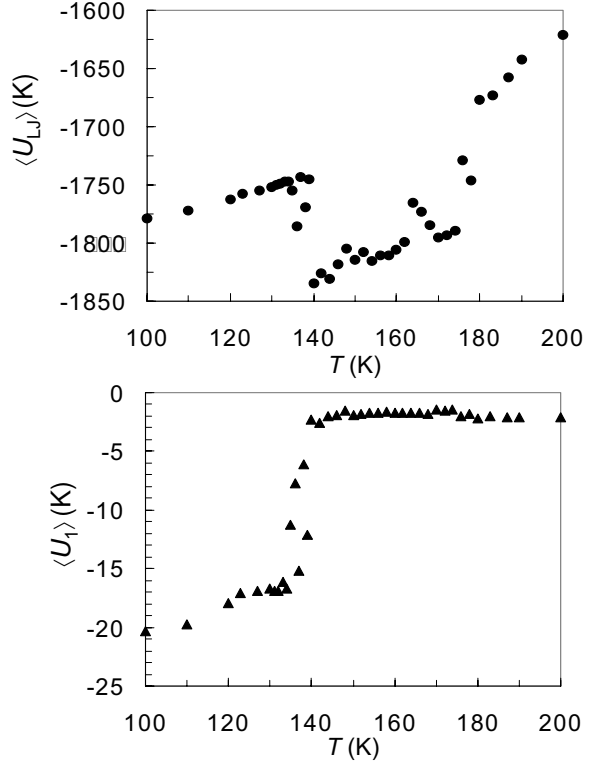}
\vspace{-.5cm}
\end{center}
\caption{\label{fig:ulj_u1} 
Thermodynamic information. Ensemble averages per molecule of the Lennard-Jones
interaction energy $\langle U_{LJ} \rangle$ and the
lateral corrugation in the adsorbate-graphite interaction 
in Steele's expansion $\langle U_{1} \rangle$. Note the sharp features
of $\langle U_{LJ} \rangle$ in both transitions, and the sharp
decrease in magnitude of $\langle U_{1} \rangle$ at the
commensurate-incomensurate/herringbone-nematic transition.} 
\end{figure}

In addition to structural indicators shown before (the various order
parameters), thermodynamic quantities are helpful in tracking phase
transitions.  Figure \ref{fig:ulj_u1} shows the temperature dependence
of the average Lennard-Jones interaction energy $\langle U_{LJ}
\rangle$ [Eq.\ (\ref{eq:lj})] and of the average corrugation potential
energy  $\langle U_{1} \rangle$ [Eqs.\ (\ref{eq:steele1}) and
(\ref{eq:steele2})].  The former is useful in delineating the system's
structural behavior irrespective of the substrate's potential, whereas
the latter is a good indicator of atomic order of the substrate and
therefore complements the molecular information provided by
$OP_{comm}$ (see Sec.\ \ref{sec:OPs}). 

The average interaction energy is very closely related to how close
together the hexane molecules are [in fact, $\langle U_{LJ} \rangle
\propto \int \! u_{LJ}(r) g(r) \, r dr$, where $g(r)$ is the atomic
pair correlation function (see Sec.\ \ref{sec:correlations})].  The
large increase in the magnitude at $T_1 = 138 \pm 2$ K is a clear
indication that molecules in the incomensurate nematic phase have
managed to become closer to the minimum of the Lennard-Jones
interaction.  At the melting transition ($T_2 = 176 \pm 3$ K)
molecular distances again become less optimal and this average energy
rises.  Both of these features are evident in the plots of
$g_{com}(r)$, see Fig.\ \ref{fig:gmol}.  The average corrugation
potential energy  $\langle U_{1} \rangle$ drammatically decreases in
magnitude at $T_1$, a clear indication of the lack of registration of
the adsorbate film and the substrate. 


Other energy ensemble averages have interesting features, but
considerably less pronounced than those for $\langle U_{LJ} \rangle$
and $\langle U_{1} \rangle$ (typically they show changes in their
derivatives with respect to $T$, but no sudden jumps).  One conclusion
that can be drawn from the behavior of these energy averages is that
the herringbone/commensurate to nematic/incomensurate transition is
driven by a competition between the intermolecular forces with the
substrate corrugation (see Sec.\ \ref{sec:variations} below for more
discussion on this).   

The specific heat at constant area $c_A$ (not shown) exhibits some
scatter, but does present two significant sharp peaks at ca.\ 138 K
and 176 K and there is some indication that there {\em may} be a
latent heat associated with both transitions.  However, the energy
distributions (not shown) near the transitions do not show any
striking bimodality which would seem to suggest that the transition is
not first-order.   As it is usual with finite-size simulations, the
determination of the order of the transitions remains elusive. 

\begin{figure}
\begin{center}
\leavevmode
\includegraphics[width=3.in]{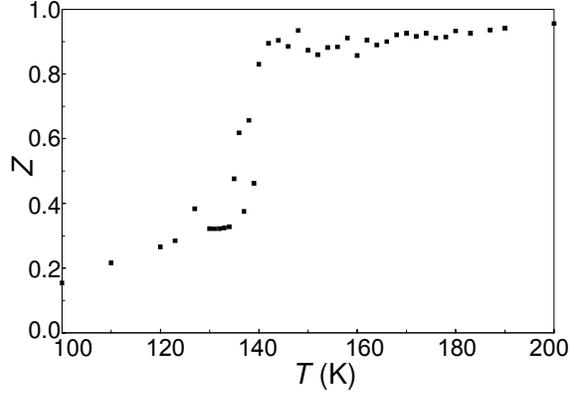}
\vspace{-.5cm}
\end{center}
\caption{\label{fig:virial} 
Relative spreading pressure $Z$ of hexane adsorbed on graphite [see Eq.\
(\protect\ref{eq:virial})]. Note that $Z \sim 0$ for $T \lesssim T_1$ 
as expected for a solid (small spreading pressure), and that $Z$ approaches 
the ideal gas value ($Z = 1$) for $T \gtrsim T_1$.   The value of $Z$ for 
the intermediate temperature region ($T_1<T<T_2$) indicates that the nematic 
is a fluid phase.} 
\end{figure}

Another quantity of interest is the {\em spreading pressure} (surface
``tension'') $\Phi$.  The virial expansion \cite{virial} 
 %
\bea
\label{eq:virial}
\Phi & \equiv & \frac{N k_B T}{A} \, Z \,, \\
Z &\equiv &  1 + \frac{1}{2Nk_BT} \left< 
    \sum_{i\neq j}^{N} \rr_{ij} \cdot \frac{\partial U}{\partial \rr_{ij}}
    \right> \nn \\
   &\simeq&  1 +  \int d^2r [g_{com}(r) - 1] \nn \,,
\eea
 %
relates it to the interactions between molecules or equivalently to the
center of mass pair correlation function $g_{com}(r)$ (see Sec.\
\ref{sec:correlations}).  Here $N/A$ is the areal density of the molecules.
The ``relative spreading pressure'' $Z$ is thus a good indicator 
of how close the equation of state 
is to that of an ideal gas ($Z \sim 1$) or that of a solid 
($Z \sim 0$).  Figure \ref{fig:virial} shows $Z$ 
for the hexane system. It is evident from our
results that below $T_1$ the system has a very small spreading pressure as
expected for a solid phase, whereas {\em both} the nematic ($T_1<T<T_2$) and
isotropic phases ($T>T_2$) are very very close to an ideal gas.  These results
are indicative of the liquid nature of the intermediate nematic phase.

\vspace{-.2cm}
\subsection{Distributions}
\label{sec:distributions}
\vspace{-.2cm}

\begin{figure}
\begin{center}
\leavevmode
\includegraphics[width=3.in]{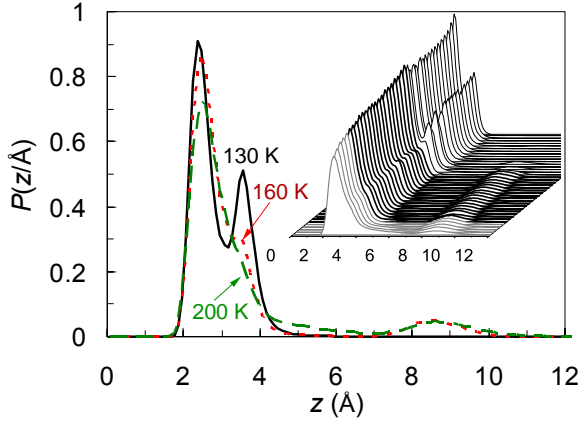}
\vspace{-.5cm}
\end{center}
\caption{\label{fig:P_z} 
Atomic height distributions $P(z)$ for the herringbone
solid at $T=130$ K (solid black line); nematic at $T=160$ K
(red dotted line); and liquid at $T=200$ K (green dashed line).  
The insert shows all results obtained from $T=100$ to 200 K back 
to front respectively.  The thin black lines represent the 
herringbone ($T \lesssim 138$ K), the thick black lines the 
nematic ($138\;{\rm K} \lesssim T \lesssim 176 \;{\rm K}$) and the gray lines 
the liquid phases ($T \gtrsim 176$ K), respectively.}
\end{figure}

\begin{figure}
\begin{center}
\leavevmode
\includegraphics[width=3.in]{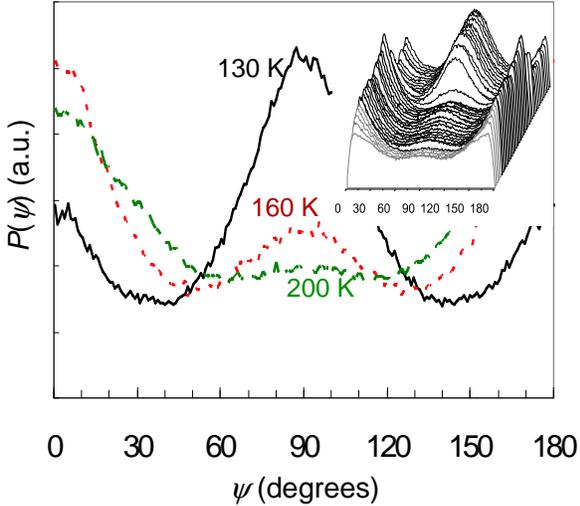}
\vspace{-.5cm}
\end{center}
\caption{\label{fig:P_psi} 
Molecular roll angle distributions $P(\psi)$ .  Please refer to Figure
\protect{\ref{fig:P_z}} for format information. Note how sharp the
change in $P(\psi)$ is at the HB to nematic transition.} 
\end{figure}

\begin{figure}
\begin{center}
\leavevmode
\includegraphics[width=3.in]{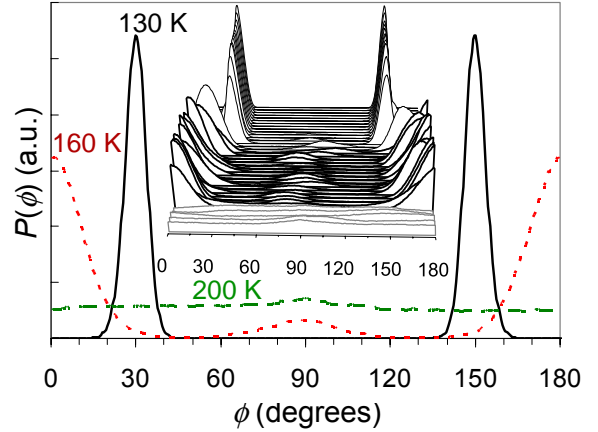}
\vspace{-.5cm}
\end{center}
\caption{\label{fig:P_phi} 
Molecular azimuthal angle distributions $P(\phi)$.  Please refer to
Figure \protect{\ref{fig:P_z}} for format information.  Note how
rapidly the position and shape of the peaks change at the HB to
nematic transition, and how fast the peaks disappear in the nematic to
fluid transition.} 
\end{figure}

\begin{figure}
\begin{center}
\leavevmode
\includegraphics[width=3.in]{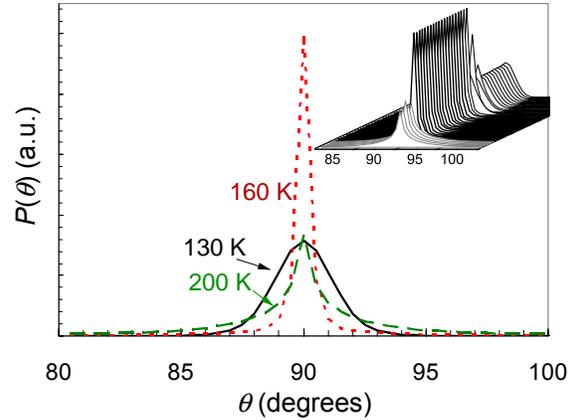}
\vspace{-.5cm}
\end{center}
\caption{\label{fig:P_theta} 
Molecular tilt angle distributions $P(\theta)$.
Please refer to Figure \protect{\ref{fig:P_z}} 
for format information.}
\end{figure}

In conjunction to the order parameters mentioned above, various
distributions can also help illustrate the properties of the system,
especially near the phase transitions, and in particular regarding the
behavior of various molecular degrees of freedom, both positional,
orientational and internal.  In Figs.\ \ref{fig:P_z}, \ref{fig:P_phi}, 
\ref{fig:P_psi}, and  \ref{fig:P_theta} we present the atomic height
$P(z)$, azimuthal angle  $P(\phi)$, internal roll-angle $P(\psi)$ and
tilt angle $P(\theta)$ distributions for various temperatures.  For
$P(z)$ we considered $0 \le z \le 12 {\;\rm \AA}$  sampling every
$\Delta z = 0.0056 {\;\rm \AA}$.  For $P(\phi)$ and $P(\psi)$ and
$P(\theta)$ we considered angles the [0$^\circ$,180$^\circ$] domain
\{[0$^\circ$,90$^\circ$] for $P(\theta)$\}, sampling every 1$^\circ$.
The azimuthal angle for a molecule, $\phi_i$ is defined to be the
angle formed between the direction of the smallest moment of inertia
projected onto the $xy$-plane and the $x$ axis.  The tilt angle
$\theta$ is defined to be the angle formed between the direction of
the smallest moment of inertia and the $z$ axis ($\theta = 90^\circ$
corresponds to a molecule laying flat on the substrate). The internal
roll angle $\psi$ is defined for a particular bond involving any three
consecutive atoms as  
\be
\label{eq:psi}
	\psi = \cos^{-1} \!\!\left\{\!
\frac{ [(\rr_{j+1} \!- \!\rr_j) \times 
(\rr_{j-1} \!- \!\rr_j)] \!\cdot \!\hat{z}} 
{|(\rr_{j+1} \!-\! \rr_j) \times (\rr_{j-1} \!- \!\rr_j)|} \!\right\} \!.
\ee
The roll-angle has a value of 0 when the plane three consecutive atoms
in a molecule is flat (parallel to the graphite basal plane) and it
equals 90$^\circ$  when the plane is perpendicular to the graphite
\cite{rollangle}. 


The atomic height distribution (Fig. \ref{fig:P_z}) shows that
essentially no molecules have been promoted to the second layer in the
low temperature phase, and only a very small fraction in the nematic
or liquid states.  The double peak stucture at low temperatures can be
interpreted easily upon consideration of the molecular roll-angle
distributions (Fig. \ref{fig:P_psi}):  as many molecules are rolled
``on their side'', raising some atoms of the hexane chain above the
main peak.   At higher temperatures molecules are flatter against the
surface and the double-peak in $P(z)$ disappears.  The azimuthal angle
distribution (Fig. \ref{fig:P_phi}) shows three distinct temperature
regions corresponding to the HB phase for $T\lesssim138$ K [$P(\phi)$
peaked at 30$^\circ$ and 130$^\circ$], a nematic for ($138\;{\rm K}
\lesssim T \lesssim 176 \;{\rm K}$) [$P(\phi)$ peaked at 0$^\circ$,
but note the presence of a small peak near 90$^\circ$ which
corresponds to the formation of domains], and a nearly isotropic fluid
for  $T \gtrsim 176$ K.  A discussion of the tilt angle distribution
(Fig.\ \ref{fig:P_theta}) is presented later (Sec.\
\ref{sec:mechanisms}), as it is linked to the mechanisms for the phase
transitions. See also the discussion on orientational order parameters
in Sec.\ \ref{sec:OPs}.

\vspace{-.1cm}
\subsection{Internal degrees of freedom}
\label{sec:internal}
\vspace{-.2cm}

\begin{figure}
\begin{center}
\leavevmode
\includegraphics[width=2.5in]{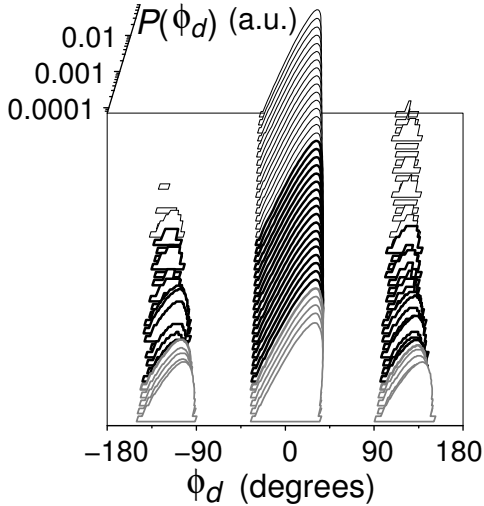}
\vspace{-.5cm}
\end{center}
\caption{\label{fig:P_phi_d} 
Dihedral angle distributions $P(\phi_d)$ for $100 \le T \le 200$ (back
to front). Please refer to Figure \protect{\ref{fig:P_z}} for format
information. } 
\end{figure}

\begin{figure}
\begin{center}
\leavevmode
\includegraphics[width=2.5in]{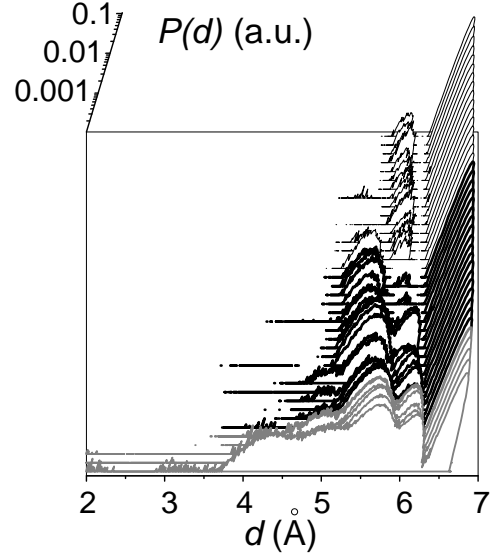}
\vspace{-.5cm}
\end{center}
\caption{\label{fig:P_d} 
End-to-end distance distributions $P(d)$ for $100 \le T \le 200$ (back
to front). Please refer to Figure \protect{\ref{fig:P_z}} for format
information. } 
\end{figure}

It is important to monitor the degree to which the phase transitions
exhibited by the system are correlated to internal degrees of
molecular freedom.  The bond-angle distribution (not shown) is a
simple gaussian corresponding exactly to the Maxwell-Boltzmann
distribution for the potential $u_{bend}$ [Eq.\ (\ref{eq:ubend})].
This is, in fact, indicative that the bond angles are unaffected by
interactions with other molecules or substrate and irrelevant to the
properties of the phases and phase transitions studied. Figures
\ref{fig:P_phi_d} and \ref{fig:P_d} show the distributions of the
dihedral angles, $P(\phi_d)$, as well as end-to end distances
(CH$_3$-CH$_3$ separations in the same molecule), $P(d)$,
respectively.   

The information on both figures is significantly correlated, as the
lower-length molecules correspond to the presence of one or more
gauche defects [the width of the main lobe in $P(d)$ corresponds to
the width of the bond-angle distribution]. We find that the number of
gauche defects is significantly smaller than for hexane molecules {\em
in vacuo} [equipartition based on $u_{tors}$, Eq.\ (\ref{eq:utors})],
mostly due to the ``flattening effect'' of the substrate potential,
but also due to the interactions with the neighbors to a lesser
degree.  
The number of gauche defects is small until close to $T_2$ and
does not present a very significant temperature dependence.  The
end-to-end probability distribution shows a slightly 
more pronounced change at
$T_1$.  Given that the fraction of molecules involved (those with
gauche defects) is so small, we find it unlikely that these internal
degrees of freedom affect significantly the phase transition at
$T_1$. For the phase transition at $T_2$ we find that gauche defects do
contribute to the footprint reduction as first suggested by Hansen {\em et al.}
\cite{flemming92,flemming93}, however that contribution does 
not appear to be as important as that from tilting (see Sec.\
\ref{sec:variations}).

\vspace{-.2cm}
\subsection{Correlations}
\label{sec:correlations}
\vspace{-.2cm}

\begin{figure}
\begin{center}
\leavevmode
\includegraphics[width=3.in]{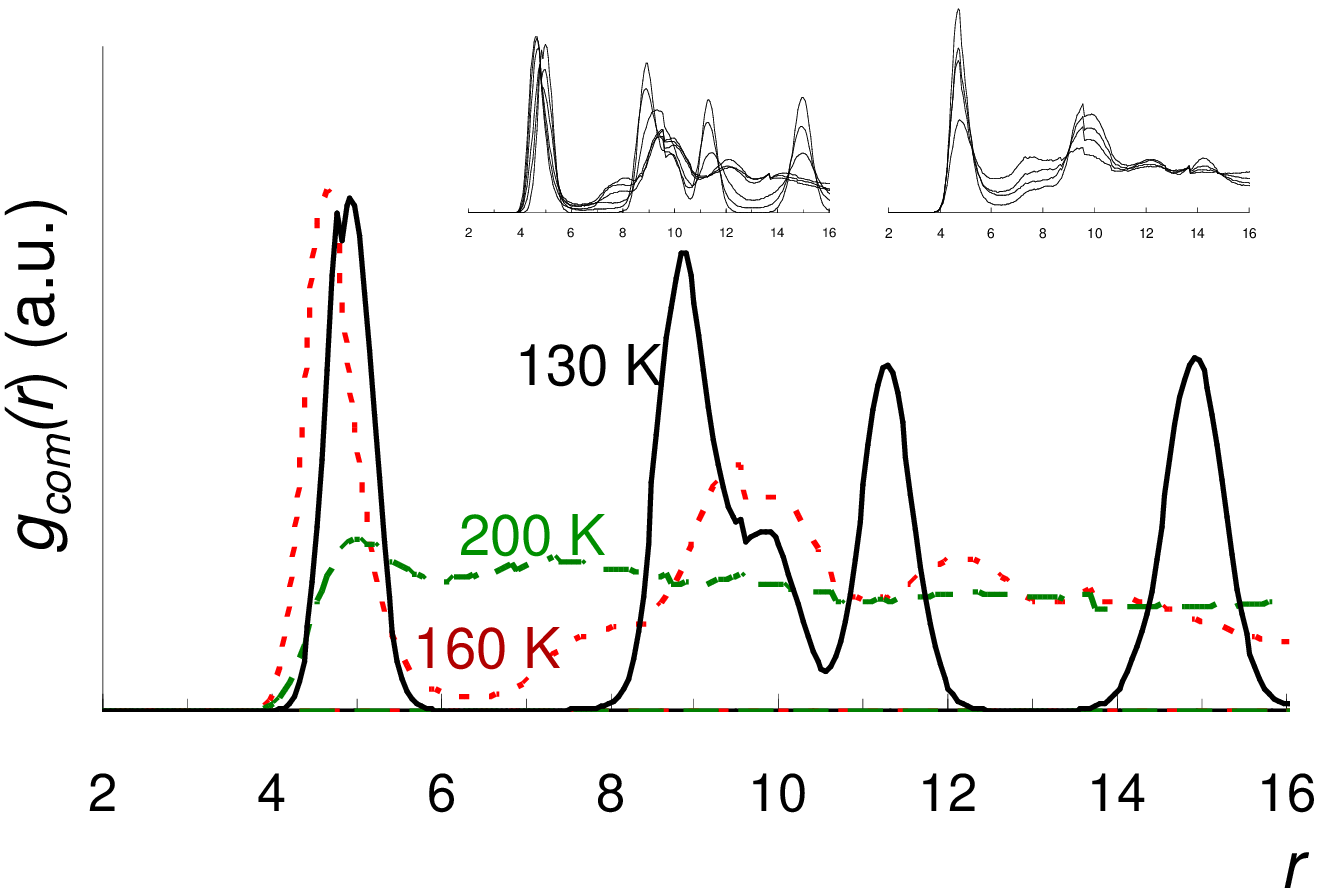}
\includegraphics[width=3.in]{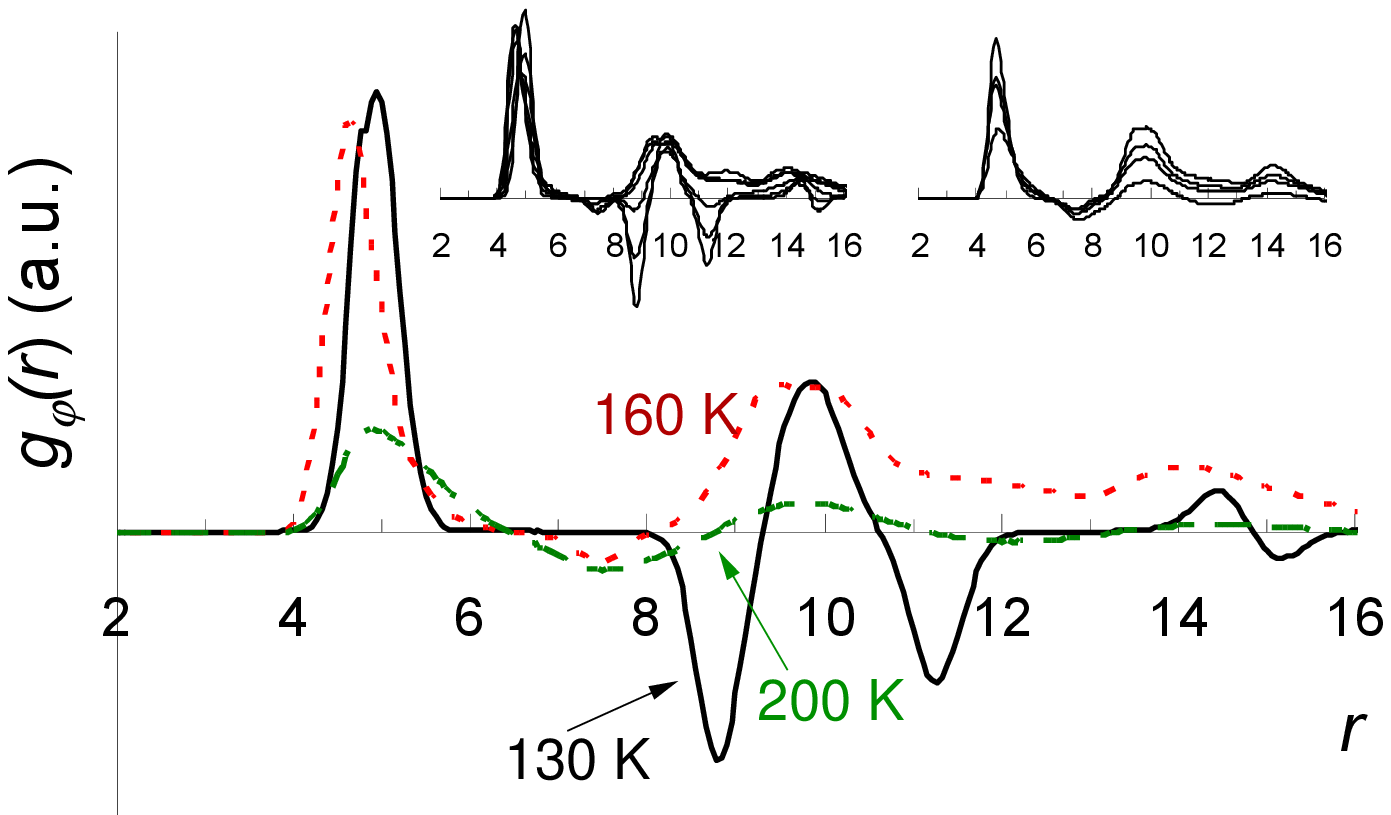}
\vspace{-.5cm}
\end{center}
\caption{\label{fig:gmol} 
Center of mass molecular pair correlation function $g_{com}(r)$ (top
panel) and molecular orientational pair correlation function
$g_\phi(r)$ for the herringbone solid at $T=130$ K (solid black line);
nematic at $T=160$ K (red dotted line); and liquid at $T=200$ K (green
dashed line). The left inserts show the transitional region between
herringbone and nematic ($130\;{\rm K} < T < 150 \;{\rm K}$), the
right inserts show the transitional  region between nematic and liquid
($170\;{\rm K} < T < 180 \;{\rm K}$).} 
\end{figure}

We have also obtained various correlation functions that permit a more
complete elucidation of the characteristics of the three phases
observed for hexane on graphite and to calculate properties that can
be directly probed experimentally [i.e.\ the static structure factor
$S(q)$, see below]. 

We start by defining the center of mass molecular pair correlation
function $g_{com}(r)$, which corresponds to the joint probability of
finding a molecule with center of mass a distance $r$ away from
another.  We also define the {\em orientational} pair correlation
function $g_\phi(r)$, which is calculated by averaging $\cos 2 (\phi_j
- \phi_i)$ ($\phi_i$ is the azimuthal angle of the $i$-th molecule)
calculated over all molecule pairs $(i,j)$ whose centers of mass have
a separation $r$. (In practice correlation functions are calculated on
a discrete sampling of the distance between the centers of mass, in
our case we used $\Delta r = 0.083$ {\AA}, also note that we use
two-dimensional normalization.) 

Figure \ref{fig:gmol} presents $g_{com}(r)$ and $g_\phi(r)$ for
various relevant temperatures.  For 130 K, both curves confirm that
the HB phase is solid [$g_{com}(r)$ has a sharp peak structure and
vanishes between peaks], and the sharp alternating signs of
$g_\phi(r)$ is indicative of the HB phase too.  At 160 K, however, the
correlation functions reveal that the mesophase is, indeed, liquid,
but with a very strong {\em orientational order} corresponding to a
nematic phase.  The evidence is quite strong towards labeling the
mesophase as a {\em nematic liquid crystal}.  At 200 K, $g_{com}(r)$
is rather smooth and $g_\phi(r)$ rapidly decays to 0, a clear
indication of a regular liquid behavior.  This behavior is consistent
with the conclusions obtained from the analysis of the spreading pressure
in Sec.\ \ref{sec:energy} (see Fig.\ \ref{fig:virial}).

\begin{figure}
\begin{center}
\leavevmode
\includegraphics[width=3.in]{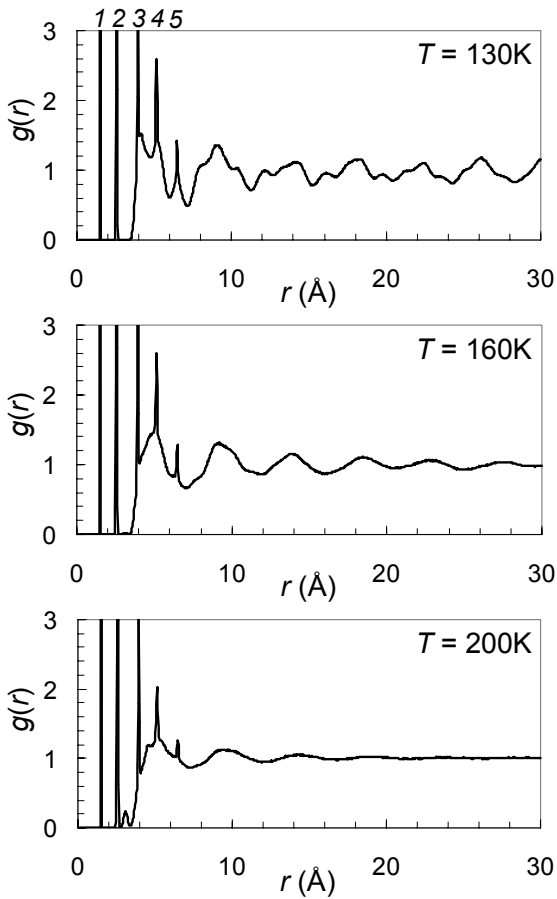}
\vspace{-.5cm}
\end{center}
\caption{\label{fig:g} 
Atomic pair correlation function $g(r)$ for the Herringbone solid (top
panel, $T=130$ K), nematic (middle panel, $T=160$ K), and liquid
(bottom panel, $T=200$ K). The sharp peaks numbered 1--5 correspond to
intramolecular pairs at distances corresponding to a hexane molecule
in its lowest configurational state: $r =$ 1.54, 2.58, 3.96, 5.16 and 6.50
{\AA}. Broader (and weaker) peaks (visible only in the lower panel) at
small distances correspond to intramolecular pairs affected by a
gauche defect: $r =$ 3.11, 4.55 and 6.05 {\AA}. The broad and
relatively smooth behavior at larger distances corresponds to pairs
belonging to different molecules.} 
\end{figure}

\begin{figure}
\begin{center}
\leavevmode
\includegraphics[width=3.in]{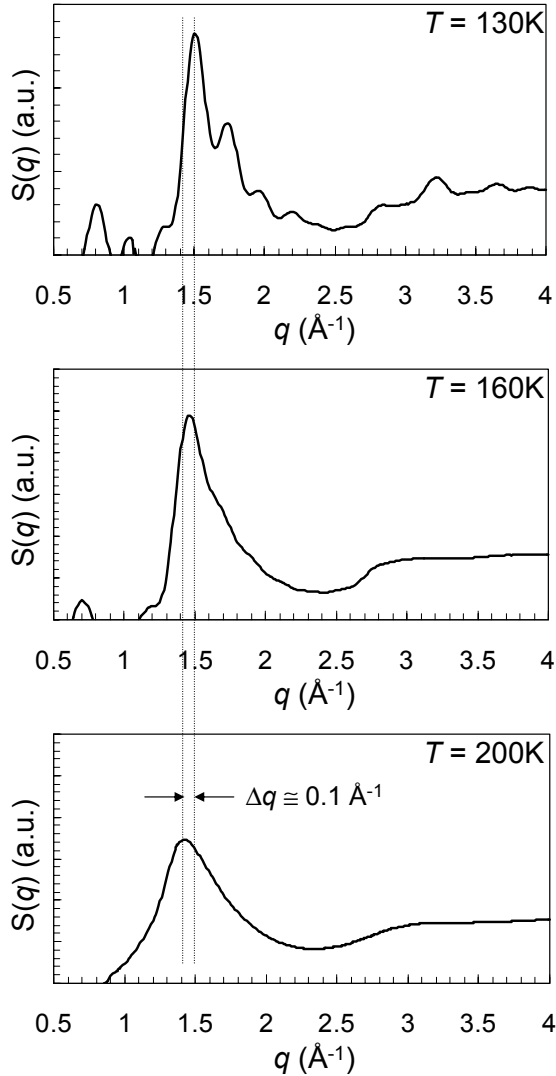}
\vspace{-.5cm}
\end{center}
\caption{\label{fig:s} 
Static structure factor $S(q)$ for the
Herringbone solid (top panel, $T=130$ K), 
nematic (middle panel, $T=160$ K), 
and liquid (bottom panel, $T=200$ K).}
\end{figure}

A second important correlation function corresponds to the {\em
atom-atom} pair correlation function $g(r)$ which is defined as the
joint probability of finding one {\em atom} at a distance $r$ from
another (again, distances were discretized with $\Delta r = 0.083$
{\AA}, and we used two-dimensional normalization).  Figure \ref{fig:g}
shows the atomic pair correlation function for three characteristic
temperatures corresponding to a HB (130 K), a nematic (160 K) and a
liquid (200 K).  In all cases, we observe very sharp peaks
corresponding to intramolecular pairs: $r =$ 1.54, 2.58, 3.96, 5.16
and 6.50 {\AA} which correspond exactly to the atomic distances in a
hexane molecule in its lowest configurational state.  Broader (and
weaker) peaks (visible only in the lower panel) at small distances
correspond to intramolecular pairs affected by a gauche defect: $r =$
3.11, 4.55 and 6.05 {\AA}. The broad and relatively smooth behavior at
larger distances corresponds to pairs belonging to different
molecules.  

From the atomic pair correlation function, one can readily obtain the
static structure factor $S(q)$ which is very important because it
contains spectral weight information in reciprocal space, and is
therefore directly connected to experimental diffraction data
\cite{newtonthesis,taub88,flemming92,flemming93,herwig97}.  Since most
experiments are performed with powder samples, we chose to calculate a
three-dimensionally averaged static structure factor: 
\be
\label{eq:sq}
   S(q) = 1 + \int_0^\infty \frac{\sin qr}{qr} [ {g(r)} - 1 ] r \, dr \,.
\ee
Figure \ref{fig:s} presents the static structure factor calculated
from the atomic pair correlation functions in Fig.\ \ref{fig:g}.  We
note a general broadening of $S(q)$ with increased temperature and
only a rather minor shift towards smaller momenta of the main peak.
The results are in qualitative agreement with experimental evidence
\cite{newtonthesis,taub88,flemming92,flemming93,herwig97}.

\subsection{Variations}
\label{sec:variations}
\vspace{-.2cm}

In order to better understand the nature of the phase transitions (in
particular the less well known HB to nematic) we ran MD simulations
where some of the interaction potentials were varied.  We considered
three interesting possibilities:  \emph{(a)} variations of the
Lennard-Jones interaction parameter $\epsilon_{\rm CH_2}$ and
$\epsilon_{\rm CH_3}$ corresponding to the pseudo atoms of the hexane
molecules [Eq.\ (\ref{eq:lj})], from 25\% to 200\% of its ``standard''
value; \emph{(b)} variations of the corrugation potential strength
in Steele's expansion [Eqs.\ (\ref{eq:steele1}) and
(\ref{eq:steele2})], from 25\% to 1,000\% of the commonly accepted
strength. ; and \emph{(c)} variations of the dihedral potential strength 
$u_{tors}(\phi_d)$ [Eq.\ (\ref{eq:utors})] from its usual strength (Table
 \ref{tabl:bond}) to ten times higher thus effectively eliminating gauche
 defects.

We find that for the first type of variations [item (a) above], both
$T_1$ (the HB to nematic transition) and $T_2$ (the nematic to
isotropic liquid transition) scale roughly like $\epsilon_{\rm
CH_n}^{1/2}$ with a correlation $R^2 \simeq 0.999$.  As $T_2-T_1$
increases for larger $\epsilon_{\rm CH_n}$ the nematic has a better
chance of ``expressing itself'' and the maximum value of $OP_{nem}$
increases too.  This simple scaling indicates that the substrate-adatom
interaction [mostly through the $E_{0i}(z_i)$ term---see Eq.\
(\ref{eq:steele1}), which holds the molecules stronger to the
surface stifling tilting] is dominant in setting the temperature
scales [since Lorentz-Bertholt rules, Eq.\ (\ref{eq:lb}), makes this
interaction scale like $\epsilon_{\rm CH_n}^{1/2}$].  The reason the
holding potential influences the phase transitions directly can be
attributed to its stiffiling effect on tilting and the formation of
gauche defects.

In the second type of variations [item (b) above], we find that $T_1$
increases with the corrugation energy, whereas $T_2$ is almost
insensitive to it.  As $T_1$ increases, the nematic phase becomes
``weaker'', eventually disappearing for a corrugation ca.\ 5 times the
standard value, and we see a direct transition between the HB and the
isotropic fluid.   These observations demonstrate that the corrugation
is irrelevant for the nematic to isotropic melting (as expected, since
this transition occurs between two incommensurate phases).
Furthermore, the increase of $T_1$ is an indication that, in addition
to the effect of holding potential described above, a competition
between the corrugation potential (which favors a commensurate phase)
and the interactions between molecules (which favors some {\em other}
characteristic intermolecular spacing) is important.   

The third type of variation [item (c) above] is useful to characterize which
degrees of freedom contribute to the footprint reduction mechanism
that creates in-plane space that facilitates the phase transitions.
Since the transition  at $T_1$ happens when the number of gauche
defects is minuscule, it is  evident that they are not important
there.  For the phase transition at $T_2$ we found (Sec.\
\ref{sec:results}) a significant increase in tilting {\em and} gauche 
defects as the system achieves its high temperature phase.  Our variation of 
$u_{tors}(\phi_d)$, which essentially eliminates gauche defects, resulted in an
increase of $T_2$ by ca.\ 20 K.  This clearly indicates that gauche defects
are important for the footprint reduction as first suggested by {\em et al.}
\cite{flemming92,flemming93}, however tilting appears to be more important.

\vspace{-.1cm}
\section{Discussion}
\label{sec:discussion}
\vspace{-.1cm}

\subsection{Herringbone commensurate solid to nematic liquid crystal
	 	transition at ${\bm{T_1 = 138 \pm 2}}$ K}
\vspace{-.2cm}

The low temperature ($T\lesssim138$ K) phase of the hexane on graphite
system is a commensurate herringbone solid, as evidenced by the
structure shown in Fig.\ \ref{fig:topview}, the values of $OP_{herr}$,
$OP_{comm}$ and $OP_{nem}$ in Figure \ref{fig:OPs}, and the two
distinct peaks in the azimuthal angle distribution $P(\phi)$ at
30$^\circ$ and 150$^\circ$ in Fig.\ \ref{fig:P_phi}. 
The spreading pressure of this solid system is very small (low temperature 
region of Fig.\ \ref{fig:virial}). Since the the monolayer is complete the
system creates in-plane room by the molecules spending a significant amount 
of their time rolled on their
side, as seen in the large peak in the roll angle distributions
$P(\psi)$ in Fig.\ \ref{fig:P_psi} as well as the double-peaked atomic
height distribution $P(z)$, see Fig.\ \ref{fig:P_z}.  

As the temperature of the system is increased, the usual signatures of
increasing thermal fluctuations appear, up until $T_1=138 \pm 2$ K,
where the system undergoes a phase change. Together, the structural
order parameters in Figure \ref{fig:OPs} indicate that the system
becomes incommensurate at the same time it loses its herringbone
orientational ordering.  In fact, this is also confirmed by the sharp
decrease in magnitude in the average Steele corrugation energy
$\langle U_1 \rangle$ (Fig.\ \ref{fig:ulj_u1}, bottom panel) indicates
that, even for atomic degrees of freedom the system is incommensurate.
The center of mass pair correlation function $g_{com}(r)$ (Fig.\
\ref{fig:gmol}, top panel) indicates that the new phase is fluid, as
evidenced by the fact that $g_{com}(r)$ has relatively smooth
features, and that the molecules are slightly closer to one another,
which also leads to an increase in the magnitude of the interaction
energy $\langle U_{LJ} \rangle$ (Fig.\ \ref{fig:ulj_u1}, top panel).
The liquid nature of this phase is also evidenced by the relatively
large spreading pressure (Fig.\ \ref{fig:virial}). 
As the solid-nematic transition proceeds, the tilt angles of the
molecules fluctuate creating a footprint reduction that facilitates
the phase transition (yet tilting is once again reduced when the new
phase is reached), as seen by the sharp peak in $OP_{tilt}$ at ca.\
138 K (Fig.\ \ref{fig:OPs}, bottom panel). 

Inspection of the snapshots in Fig.\ \ref{fig:topview}, plus more
quantitative considertions based on $OP_{nem}$ and $P(\phi)$ further
indicates that the positional order lost in this phase transition
occurs simultaneously with a new 
orientational order where all molecules have a common director. The
molecules experience, in fact, a relatively long range orientational
order (see 160 K curve, Fig.\ \ref{fig:gmol}, bottom panel). We
believe that this phase has all the characteristics of a two
dimensional {\em nematic liquid crystal}. 
Analysis of the probability distributions for the bond and dihedral
angles, and for the molecular end-to-end distance indicate that the
internal degrees of freedom do not cooperate significantly with the HB
to nematic phase transition.   
In fact, it seems that they hardly participate at all in the relevant
processes in the temperature range studied for a monolayer of hexane
adsorbed on graphite.  It should be noted that number of gauche
defects is significantly smaller than what would be predicted {\em in
vacuo}, which is due to a large degree to 
the substrate interaction, and to a lesser degree to the interaction
between molecules which has an aligning effect. 


\vspace{-.2cm}
\subsection{Nematic to isotropic liquid phase transition 
	at ${\bm{T_2 = 176 \pm 3}}$ K}
\vspace{-.3cm}
	
As the temperature is raised, the nematic persists up until $T_2 = 176
\pm 3$ K where another orientational phase transition occurs, this
time to an isotropic fluid.  The phase transition is characterized by
a sudden drop in the nematic order parameter $OP_{nem}$ (Fig.\
\ref{fig:OPs}, center panel) and the disappearance of any directional
preference in the azimuthal angle distribution $P(\phi)$ (Fig.\
\ref{fig:P_phi}, see also the random orientation in the 200 K snapshot
in Fig.\ \ref{fig:topview}). In addition, the orientational pair
correlation function $g_\phi(r)$ shows that angular correlations
between molecules decay extremely rapidly with distance in the high
temperature phase (see 200 K curve, Fig.\ \ref{fig:gmol}, bottom
panel). The Lennard-Jones interaction increases sharply (which is
related to the increase in the typical intermolecular distances), and
there is a strong peak at at $T_2$ in the specific heat $c_A$.  

In the liquid, the tilt angle behavior shown in the bottom panel of Fig.\
\ref{fig:OPs} reveals that the molecular axes are much freer to
fluctuate than in the nematic and this may cause a significant reduction 
in the molecular footprint (as discussed in Sec.\ \ref{sec:variations}, gauche
defects also contribute to this footprint reduction, though 
tilting appears to be more important).
The dihedral angle and end-to-end distance distributions in Figs.\
\ref{fig:P_phi_d} and \ref{fig:P_d} show that, in the liquid, gauche
and other dihedral defects are promoted more readily with increasing
temperature, but, as with the solid-to-nematic phase transition, no
new or abrupt changes in the internal degrees of molecular freedom
accompany melting. 

\vspace{-.2cm}
\subsection{Information related to scattering}
\vspace{-.3cm}

The atomic pair correlation functions in Fig.\ \ref{fig:g} are a
combination of intra- and intermolecular pair terms and hence are rich
in information about how the internal degrees of freedom behave as
well as the bulk structure of the system's phase.  The five extremely
sharp peaks correspond to atom-atom distances within the molecules,
and correspond precisely to the intramolecular atomic distances in the
ground state of hexane.  Careful inspection of the $g(r)$ curves at
high temperatures reveals the emergence of smaller more diffuse peaks,
which again precisely correspond to the intramolecular atomic
distances, this time in presence of one gauche defect.  When the curve
$g(r)$ starts to elevate off the horizontal axis, the contribution
from atom pairs between different molecules begins and the decreasing
structural order with increasing temperature is evident. The static
structure factor $S(q)$ shown in Fig.\ \ref{fig:s} helps further
understand the behavior of $g(r)$.  We note a general broadening of
$S(q)$ with increased temperature and only a rather minor shift
towards smaller momenta of the main peak.  The results are in
qualitative agreement with experimental evidence
\cite{newtonthesis,taub88,flemming92,flemming93,herwig97}. 

\vspace{-.5cm}
\subsection{Mechanisms}
\label{sec:mechanisms}
\vspace{-.4cm}


As in previous computer simulations
\cite{velasco95,peters96a,peters96b}, our work reproduces general
features of the real system as explored by experiment, and quite
accurately determines the melting temperature $T_2$
\cite{krim85,newtonthesis,taub88,flemming92,flemming93,herwig97}. Our
findings of an intermediate {\em nematic} liquid crystalline phase
between $T_1$ and $T_2$ are roughly in agreement with previous
simulations by Peters and Tildesley \cite{peters96a}.  However, the
existence of this intermediate phase has not been confirmed
experimentally, which suggests that larger computer simulations would
be of interest in determining the presence of highly complicated
domain behavior, or even dynamic domains that smaller simulations
simply cannot capture.   

A unique benefit of computer simulations is that they allow careful
inspection of the mechanism for various phase transitions when enough
data is present. In the work presented here, we see that in the solid
herringbone phase the molecules spend much time rolled so that their
backbones are ziz-zagged, or rough relative to the graphite
substrate (see Fig.\ \ref{fig:P_psi}).  The solid to nematic transition 
involves negligible amount of gauche defects (Fig.\ \ref{fig:P_phi_d}), 
instead showing pronounced molecular tilt fluctuations 
(Fig.\ \ref{fig:OPs}, bottom panel)
in concert with the molecules' rolling on their sides. The 
resulting nematic phase still has an average density
of unity but the molecules are closer to each other (in the direction
perpendicular to the molecules), as evidenced by the shift in the first 
peak of the center of mass pair distribution function $g_{com}(r)$ (Fig.\ 
\ref{fig:gmol}) and the sudden increase in magnitude of the Lennard-Jones
interaction $\langle U_{LJ}\rangle$ (Fig.\ \ref{fig:ulj_u1}, top panel).
As a result, the tilt-angle distribution (Fig. \ref{fig:P_theta}) takes on a
fundamentally different shape somewhat more sharply peaked that the 
gaussian-like distributions seen in the solid. 
The importance of the molecule-molecule interaction in the mesophase is
underscored by the fact that the tilt angle distributions do not
considerably broaden in the nematic.  

At a higher temperature the
system melts onto an isotropic phase. Although no sudden increase in
tilt fluctuation accompanies melting, the tilt angle distributions
begin to show thermal broadening, accounting for the increased slope
of $OP_{tilt}$ in the liquid (Fig.\ \ref{fig:OPs}, bottom panel).  
A possible ``signature'' of the nematic is that the 
tilt angles are locked and that it is a ``tilt-locking''
mechanism that preceeds the mesophase.  The first simulations
completed on the system \cite{flemming92,flemming93} concluded that,
at melting, proliferation of dihedral (gauche) defects in the
adsorbate cause the projected presence of the molecules (their
footprint) to decrease, thus creating more in-plane room for
center-of-mass molecular fluctuations and thus allowing melting to
ensue. In fact, freezing of the dihedral degrees of freedom resulted
in a considerable increase in the system's melting temperature.  Later
work by the same authors \cite{herwig97} and others
\cite{velasco95,peters96a,peters96b} suggests that, when the
molecule-substrate interaction strength is decreased to 
more realistic values, guche defects become unimportant for melting. 
Our work suggests that the creation of in-plane free space is
indeed important for all phase transitions in the system, but that the 
``footprint reduction'' happens exclusively through tilting (and moderate
stacking) at the solid--nematic transition, where the internal molecular 
degrees of freedom are unimportant.  The ``footprint reduction'' for the 
nematic--isotropic transition is, indeed, caused by both gauche defects and 
tilting, as both the probability of finding significant dihedral angles 
{\em and} tilting start rising significantly near $T_2$ (see bottom panel
 of Fig.\ \ref{fig:OPs} and Fig.\ \ref{fig:P_phi_d}).  
  
The simulations conducted on the variation of various Lennard-Jones
interaction parameters agree with the above assessments because when
only the lateral graphite corrugation well depth is varied [type (b)
as defined in Sec.\ \ref{sec:variations}], $T_1$ increases slightly 
and $T_2$ remains unchanged. However when the potential parameters
for the pseudoatoms themselves are varied [type (a)], the interaction
of the adatoms with the substrate also increases, including a stronger
vertical force component. Thus, tilting is stifled and $T_1$ and 
$T_2$ are significantly affected. Since we already know that an
increase in lateral forces alone cause relatively small changes in
$T_1$, such results underscore the importance of tilting prior to the
nematic.  Our elimination of gauche defects by varying the torsional potential
[variation type (c)] resulted in a moderate increase of the transition
temperature $T_2$, from which we conclude that both tilting and gauche
defects are significant, albeit it seems that the former are more
important. In essence, we agree qualitatively with earlier work's
findings  that ``footprint reduction'' is important in this system,
only in our work  the prevalent mechanism is tilting.


\vspace{-.1cm}
\subsection{Final Remarks}
\vspace{-.2cm}

Our simulations indicate that the HB to
nematic transition is due to a competition between the
pro-commensurate interaction between adsorbate molecules and the
corrugation substrate potential and the intermolecular interactions
which favors intermolecular distances different from the required by
the substrate.  The latter prevail at ca. 138 K, leading to a nematic
liquid crystalline state.  If a stronger attraction to the substrate
is used in our similations (Sec.\ \ref{sec:variations}) the stifling
of mechanisms that reduce the molecular footprint makes the phase
transition occur at a higher temperature.  An increase in the
corrugation potential has a similar effect. 
The nematic to isotropic transition is, as expected, rather
insensitive to the corrugation potential, but remains very sensitive
to the main part of the substrate-adatom potential.

We have observed somewhat contradictory evidence in terms of trying to
identify the order of the phase transitions.  Whereas the specific
heat curves seem to indicate that there may be a latent heat in both
cases, the lack of bimodality in the energy distributions would
indicate otherwise.  The inconclusiveness is not, however, surprising
given the finite size of the system simulated. 

We have also performed MD simulations with ca.\ $2 \times$ and $4
\times$ the number of molecules discussed throughout this paper.  We
find that the HB to nematic phase transition temperature $T_1$ is only
weakly dependent on system size (but the relatively poor statistics
possible with the larger sizes make it difficult to extrapolate to the
thermodynamic limit).  The nematic to isotropic phase transition
temperature $T_2$ is insensitive to system size for the sizes
considered.

\vspace{-.3cm}
\acknowledgments
\vspace{-.2cm}

The authors are indebted to Haskell Taub, Flemming Hansen, G\"uenther
Peters, Peter Pfeifer, Cintia Lapilli and James MacCollough for
enriching discussions. Acknowledgement is made to the Donors of the
Petroleum Research Fund, administered by the American Chemical
Society, for support of this research.  CW acknowledges support from
the University of Missouri Research Board and the University of
Missouri Research Council.  MR and CP acknowledge support by the
University of Northern Iowa Faculty Summer Research Fellowship and
University of Northern Iowa Summer Undergraduate Research Fellowship
respectively.

\vspace{-.0cm}


\begin{thebibliography}{99}
\bibitem{directors}
	Note that since hexane is not polar, we are dealing with {\em
	directors} rather than vectors, hence the $[0,\pi]$ range for
	$\phi_i$, similar arguments also apply to other molecular directions 
	($\theta_i \in [0,\pi/2]$, $\psi \in [0,\pi]$).
\bibitem{rollangle}
	The internal roll angle is related to the rolling angle in previous
	work \cite{peters96a,peters96b}, with the exception that it gives
	information about the rotation of the planes that each bond angle lies
	in as opposed to overall molecular orientation.
\end{thebibliography}
\end{document}